\documentclass[onecolumn,preprintnumbers,amsmath,amssymb]{revtex4}
\input{epsf}

\newcommand{\treh}{T_{\rm R}}
\newcommand\gravitino{\widetilde{G}}    
\newcommand\mgravitino{m_{\gravitino}}
\newcommand\tanb{\tan\beta}
\newcommand\sgn{{\rm sgn}}

\newcommand\tev{\,\mbox{TeV}}
\newcommand\gev{\,\mbox{GeV}}

\newcommand\kev{\,\mbox{keV}}

\newcommand\gluino{\tilde g}
\newcommand\mgluino{m_{\gluino}}

\newcommand{\mhalf}{m_{1/2}}      \newcommand{\mzero}{m_0}
\newcommand\mchi{m_{\chi}}              
\newcommand{\stauone}{{\tilde \tau}_1}   \newcommand\mstauone{m_{\stauone}}
   
\newcommand{\stauright}{{\tilde \tau}_R} 
      \newcommand{\mz}{m_{Z}}

\newcommand{\be}{\begin{equation}}
\newcommand{\ee}{\end{equation}}
\newcommand{\bea}{\begin{eqnarray}}
\newcommand{\eea}{\end{eqnarray}}
\newcommand{\bef}{\begin{figure}}
\newcommand{\eef}{\end{figure}}

\newcommand{\simge}{\,{}^>_{\sim}\,}
\newcommand{\simle}{\,{}^<_{\sim}\,}

\def\h#1{$^{#1}$H}
\def\he#1{$^{#1}$He}
\def\li#1{$^{#1}$Li}
\def\be#1{$^{#1}$Be}

\def\eps@scaling{0.96}

\def\showone#1{
  \centering
  \leavevmode
  \epsfxsize=\eps@scaling\linewidth
  \epsfbox{#1.eps}
}

\def\epstwo@scaling{0.48}

\def\showtwo#1#2{
  \centering
  \leavevmode
  \epsfxsize=\epstwo@scaling\linewidth
  \epsfbox{#1.eps} \hfil
  \epsfxsize=\epstwo@scaling\linewidth
  \epsfbox{#2.eps}
}

\begin{document}

\title{Solving the Cosmic Lithium Problems with Gravitino Dark Matter in the
CMSSM}
\author{Karsten Jedamzik}
\address{Laboratoire de Physique Th\'eorique et Astroparticules,
CNRS UMR 5825,\\
Universit\'e Montpellier II, F-34095 Montpellier Cedex 5, France}

\author{Ki-Young Choi}
\address{Department of Physics and Astronomy, University of Sheffield, 
Sheffield, S3 7RH, UK}

\author{Leszek Roszkowski}
\address{Department of Physics and Astronomy,
University of Sheffield, Sheffield, S3 7RH, UK}

\author{Roberto Ruiz de Austri}
\address{Departamento de F\'{\i}sica Te\'{o}rica C-XI
 and Instituto de F\'{\i}sica Te\'{o}rica C-XVI,
 Universidad Aut\'{o}noma de Madrid, Cantoblanco,
 28049 Madrid, Spain}

\begin{abstract}
Standard Big Bang nucleosynthesis at baryonic density as inferred by WMAP
implies a primordial \li7 abundance factor of two to three larger than that
inferred by observations of low--metallicity halo stars. Recent observations
of \li6 in halo stars suggest a pre--galactic origin of this
isotope, which is exceedingly difficult to explain by putative high redshift
cosmic ray populations. We investigate if one or both of these lithium
problems
may be solved by late decaying relic particles 
in the Constrained Minimal
Supersymmetric Standard Model (CMSSM) coupled to gravity. 
Assuming that the gravitino is the lightest supersymmetric particle
(LSP) we find that in large parts of the CMSSM parameter space both of
these problems can be potentially solved. In particular, we find
solutions where both lithium problems may be solved simultaneously. 
These entail the hadronic decay of relic ${\cal O}(1\tev)$ staus into
${\cal O}(100\gev)$ gravitinos at $\sim1000$ sec after the Big Bang,
as proposed by one of us before~\cite{Jedamzik:2004er}. Moreover, the produced gravitinos 
naturally contribute a large fraction, or all, to the by WMAP required
dark matter density. A prediction of this model is the dark matter to
be lukewarm with free--streaming lengths of the order of a $4\kev$
early freezing--out relic particle. Such lukewarm gravitino dark
matter may have beneficial effects to the formation of galactic
structures and may be constrained by the reionisation history of the
Universe. The current version of the paper presents results for both cases
(a) when catalytic nuclear reations are included and (b) when they are 
neglected. 
\end{abstract}



\maketitle
\section{Introduction}\label{sect:intro}


Determining the nature of the cosmological dark matter is one of the
main outstanding challenges in modern cosmology. Compelling
candidates for the cold dark matter may be in the form of new neutral
stable particles arising in phenomenologically successful
supersymmetric extensions of the Standard Model of particle
physics. These include the neutralino~\cite{jkg}, a mixture of the
supersymmetric partners to the neutral Higgs, SU(2) $W$, and U(1) $B$
bosons, the axion and its supersymmetric partner the
axino~\cite{cdm-axino} and the
gravitino~\cite{gravitinoproduction,MMY93} (the supersymmetric partner
of the graviton). Whereas the bulk of studies has been devoted to the
neutralino, the case of gravitino dark matter has recently attracted
much attention. Studies of gravitino dark matter have been performed
either in the context of supersymmetry breaking in a hidden sector
being communicated to the visible sector by gravitational
interactions~\cite{grav-recent1,grav-recent2,eoss-grav,Wang:2004ib,rrc04,Cerdeno:2005eu}
or by gauge interactions~\cite{GMSB}.  Gravitinos may be produced in
the early Universe by (at least) two mechanism: (i) scattering of
thermal radiation at the highest temperatures of the early Universe, 
hereafter referred to as thermal production (TP) and (ii) freezeout 
and decay of 
the next--to--lightest supersymmetric particle (NLSP) to gravitinos,
hereafter referred to as non--thermal production (NTP).  Whereas the
gravitino yield in TP depends on the reheating temperature $\treh$,
the gravitino abundance due to NTP may in principle come by itself,
and independently of $\treh$~\cite{remark1}, very close to the range 
inferred for dark matter density from cosmological observations.
In the case of the gravitino LSP,
NLSPs typically decay during or after the epoch of
Big Bang nucleosynthesis (BBN), unless the gravitinos are rather light, 
thereby potentially disrupting light--element
yields~\cite{NLSPdecay,MMY93}. 
This has been often taken to disfavor for heavy
gravitino dark matter. However, though stringently constrained by BBN,
recent careful study shows that much viable gravitino dark matter parameter
space remains~\cite{grav-recent1,grav-recent2,rrc04,Cerdeno:2005eu}.

The epoch of BBN has long been known to synthesize the bulk of the
\he4 and D, as well as good fractions of the \li7 and \he3 in the
presently observed Universe. Paramount to the realization of this fact was
also the discovery of the \li7--''Spite'' plateau in 1982~\cite{spite}, 
in particular, the
observation of constant \li7/H abundances in low--metallicity Pop~II stars
over a wide range in metallicity. This indicated a pre--galactic
origin of \li7, as other sources (i.e. galactic
cosmic ray nucleosynthesis) predicted a rise of \li7 with metallicity.
Current observational estimates~\cite{li7,asplund,lambert} 
of \li7/H range between $1.10\times 10^{-10}$ 
and $2.34\times 10^{-10}$,
with differences mostly depending on which effective stellar temperature 
calibration for the Pop~II stars is used. With the accurate
estimate of the baryonic density by WMAP, i.e. 
$\Omega_bh^2\approx 0.022-0.023$,  
it was possible to predict the primordial \li7/H abundance 
of \li7/H $\approx 3.82-4.9\times 10^{-10}$~\cite{SBBN} 
within the framework of
a standard BBN (SBBN) scenario.
It is apparent that this predicted abundance is
a factor of $2-3$ larger than that observed. 

It is conceivable that \li7 in the atmospheres of Pop~II stars has been
transported down beyond the base of the convective zone of 
the stars, and thereby
depleted by nuclear burning (i.e. \li7$(p,\alpha )$\he4). Though
standard stellar models may not account for this \li7 depletion within 
the near--turnoff, low--metallicity stars of the Spite plateau, effects
not included in standard models, such as rotation, atomic diffusion, or
gravity waves 
could potentially change this conclusion. Nevertheless, rotationally induced
depletion of \li7 by a factor of $2-3$
predicts a spread in stellar \li7 abundances~\cite{Pins:01}, 
not observed by any group
(see, however, Ref.~\cite{MerchantBoesgaard:2005mi}).
This conclusion could be potentially changed when either internal stellar
gravity waves~\cite{TC04} or magnetic fields
are considered in conjunction with stellar rotation. 
Atomic diffusion, a process required to understand
the structure of the Sun by helioseismology, predicts a slope in the
plateau (as a function of stellar temperature)~\cite{MFB84} 
which is not observed in the
data (see, however, also Ref.~\cite{Sala:01}). 
Only when atomic diffusion is coupled with an ad hoc and fine--tuned
weak turbulence in the radiative zone of the star, may a depletion of
the required factor of $2-3$ result~\cite{Richard}. However, even in this case 
a dispersion of the \li7 data is likely to emerge. 
It is thus not impossible that depletion of \li7 on
the Spite plateau has occurred.
Nevertheless, \li7 depletion in
low--metallicity stars is a well-- and long-- studied possibility,  
and even after about 20 years of efforts no consistent and well-motivated 
scenario including a factor of $\simge 2-3$ depletion has yet emerged. 

Standard BBN leads to the synthesis of a \li6 abundance of \li6/H 
at the level of $10^{-14}-10^{-13}$, orders
of magnitudes below what is observable by current technology. Observed 
\li6 abundances in the Sun, galactic disc--, and halo-- stars are thus 
traditionally not believed to be of primordial origin, 
but rather due to galactic cosmic
ray nucleosynthesis via supernovae produced energetic $p$, $\alpha$ or $CNO$
inducing spallation $p + CNO\to {\rm Li\,Be\,B}$ or fusion 
$\alpha + \alpha\to {\rm Li}$
reactions in the gas of the interstellar medium. Here \li6 is produced
along with \li7,\be9, $^{10}$B, and $^{11}$B. Typical production ratios of
\li6 /\be9 $\sim 5-10$ in galactic cosmic ray nucleosynthesis are consistent
with those observed in the Sun, but not with those \li6/\be9 $\approx 40-80$
ratios observed in low--metallicity $[Z]\sim -2$ Pop~II halo stars. 
It is thus clear 
that, at the very least, the composition of galactic cosmic rays at higher redshift
has to be strongly modified in order to account for the observations.
Nevertheless, a \li6/H abundance ratio of $\sim 10^{-11}$ 
at $[Z]\sim -2$ was concluded to be only with difficulty, or not at all,
synthesized by galactic cosmic rays, an argument based on the energetics
of supernovae generated cosmic rays~\cite{Rama:99}.

It had therefore been speculated that the \li6 abundance in Pop~II stars
may originate from the early Universe, via the electromagnetic decay of
a relic particle, 
such as the gravitino, inducing the non--thermal nuclear
reaction sequence of \he4$(\gamma ,p)$\h3 photodisintegration with the
resulting energetic \h3 (and \he3) further fusing on \he4 
to form \li6~\cite{Jeda:00,remark2}.  
Here it was found that the synthesis of \li6 was efficient enough to produce
the Pop~II abundance without disturbing the other light isotopes or the
Planck spectrum of the cosmic microwave background (CMBR) at the
observable level. With the advent of the first fully coupled calculations
of thermal nuclear reactions and cascade nucleosynthesis 
{\it during} BBN~\cite{Jedamzik:2004er,Kawasaki:2004yh}
further solutions for the synthesis of the \li6 abundance in the early
Universe were found. They were based on either the residual hadronic
annihilation of a population of dark matter~\cite{Jedamzik:2004ip}
(e.g. neutralinos) during and
towards the end of BBN, or the hadronic decay of a relic, long--lived
particle population around $10^3$ sec after the 
Big Bang~\cite{Jedamzik:2004er} (e.g. gravitinos).
In hadronic decays or annihilations the initial photodisintegration
reaction in the non--thermal nuclear reaction sequence given above 
is replaced by \he4$(N,p)$\h3 spallation by energetic nucleons $N$ 
produced during the decay.

An intriguing further consequence of the hadronic {\it decay} of
a relic particle at $10^3$ sec is the prediction of a significant
\li7 abundance reduction
concomitant with the \li6 production. 
This occurs due to the thermal nuclear
reaction sequence \be7$(n,p)$\li7 and \li7$(p,\alpha )$\he4 induced by the
excess neutrons due to the decay. At the same time the D is increased
due to p$(n,\gamma )$D , but not as much as to exceed a conservative 
observational upper limit of D/H $\simle 5.3\times 10^{-5}$ 
on this isotope.
It was found~\cite{Jedamzik:2004er} that a relic particle 
abundance of $\Omega_Xh^2B_h\approx 1-5\times
10^{-4}$ (depending on the mass of the relic), where $B_h$ is the hadronic
branching ratio of the relic $X$, was sufficient to explain 
qualitatively and quantitatively
both, the low observed \li7 abundance and the high observed \li6
abundance. In~\cite{Jedamzik:2004er} it was further speculated that among
other possibilities, the relic could be
a supersymmetric stau of $\sim 1\tev$ 
mass decaying into a LSP gravitino of mass $50\gev$.
A clear prediction of the proposed scenario is the existence of a \li6
plateau, analogous to the \li7--Spite plateau, i.e. constant \li6 /H in 
low--metallicity stars.

Earlier suggestions of a primordial solution of the \li7 
discrepancy~\cite{grav-recent2} invoked 
an electromagnetic decay of a relic particle around $2\times 10^6$sec,
thereby photodisintegrating \be7. Synthesis of \li6 during the same
process was not considered. 
It was subsequently
shown~\cite{Ellis:2005ii} that such a scenario may not work, as either the 
observational upper limit on the primordial \he3/D~\cite{Sigl:1995kk} 
ratio is surpassed due to concomitant \he4 photodisintegration,
or a reasonable lower limit on the primordial
D /H $\simge 2.2\times 10^{-5}$ due to D photodisintegration
is violated.

Over the last months the number of preliminary detections of \li6/H 
in Pop~II stars has multiplied by a large factor. There are now around
ten claimed~\cite{asplund,6Li} detections of the \li6/\li7 
isotope ratio, with all ratios
falling in the range between \li6/\li7 $\approx 0.03 - 0.07$ 
(and with average \li6/\li7$\approx 0.042$), 
independent of stellar metallicity falling
in the range $[Z]\sim -2.75$ and $[Z]\sim -1.2$. There exist also around
ten upper limits, with all of the stars, nevertheless, consistent with
\li6/\li7 in their atmospheres on the level 
of $\simge 0.01$~\cite{private1}.
It is thus intriguing that the \li6 data indeed shows a plateau within
a large metallicity range. From a galactic cosmic ray
point of view the high \li6/\li7 abundances as reported in the lowest
$[Z]\sim -2.75$ metallicity star LP815--43 is particularly difficult to explain.
Following the announcement of results of 
these difficult observations with the VLT 
telescope, preliminary detections of more \li6/\li7 ratios by the
Subaru telescope and Ref.~\cite{christian} 
have been claimed, with, in particular, one 
star of metallicity $[Z]\sim -3.25$ seemingly showing again a similar
\li6/\li7 ratio~\cite{private2}. 
All this data points to the existence of a \li6
plateau (see below, however). 
Nevertheless, it needs to be stressed that each individual detection
of \li6/\li7 due to these difficult observations is only at the about
2--4 sigma level. In this sense the observations have to be taken as
preliminary.

Though the observations seem to indicate a \li6 plateau over a wide
range in metallicity (in particular also, when disk stars at metallicity
$[Z]\sim -0.6$~\cite{li6:disk} are included into the sample), a plateau may not 
necessarily exist if one is to believe the
claimed~\cite{richardpriv} metallicity dependence of \li6 destruction 
on the pre--main--sequence
(PMS) of the observed stars. Predicting \li6 destruction during this phase
is less certain~\cite{prof} 
than during the stellar main sequence. In fact, the prediction are not
comparing well to observations and are therefore suspect~\cite{piau}.
The predicted \li6 
destruction~\cite{richardpriv} would imply a rise of \li6 with
metallicity for stars with $[Z]\simge -2$. This indeed 
is theoretically favorable as
most groups also observe a rise of \li7 with metallicity on the
Spite--(quasi)--plateau, usually attributed to cosmic ray production
of \li7. As \li6 is produced by the same process a consistent
picture would emerge. In the absence of PMS \li6 destruction
either a fine--tuned stellar main--sequence \li6 destruction would have to
be invoked~\cite{Prantz}, or the conclusion of \li7 rising
with metallicity would be erroneous. Here the latter possibility seems
likely as the existence of a slope in the Spite-plateau is not confirmed
by all groups. In contrast, though further
observations are required, stars with metallicities 
$[Z]\simle -2$ seem currently, in any case, to be 
consistent with a low--metallicity \li6 plateau, 
even when stellar PMS effects are included. 

The new \li6 data has already prompted the first attempts
to be explained in terms of cosmic ray nucleosynthesis. In Ref.~\cite{Suz}
cosmic ray nucleosynthesis with energetic $\alpha$'s generated at the shocks
resulting in merger events during the formation of the Milky Way
were proposed to possibly generate the \li6 abundance of $\sim 10^{-11}$. 
This possibility, however, was subsequently withdrawn by the authors
due to failings on energetic grounds.
In Ref.~\cite{Rollinde:2004kz} 
a high redshift ($z\simge 10$) cosmic ray $\alpha$--burst was
postulated and claimed to account for the data but no suggestion about
the origins for these cosmic rays was made. The sources which may have
a cosmic ray fluence 
 sufficient to synthesize such large amounts of \li6 were
analysed in detail in Ref.~\cite{Prantz}. 
It was concluded that typical core collapse
supernovae, believed to be the source of the standard cosmic rays, fail by
a large factor. A similar conclusion was reached for shock--generated cosmic
rays at the formation of the Milky Way. Only two candidates seemed to fulfil
the energetic requirements, with both, nevertheless, involving fairly drastic
and observationally unconfirmed assumptions. The \li6 may have been due to
cosmic rays generated by a very early and efficiently forming 
population of 30--100 $M_{\odot}$ stars
(involving around 10\% of all baryons in the Milky Way halo), forming black
holes at the end of their lifes and ejecting only negligible amounts of 
iron. This scenario implies a large variation of the so far believed
universal Salpeter stellar mass function and would lead to a large number
of supermassive black holes in the galactic halo.
Alternatively, the energetic particles responsible for the \li6 synthesis
may be due to accretion of baryons on the black hole of the galactic center
of the Milky Way, if the black hole formed before the assembly of the
galactic halo and, if accretion on the black hole was around a factor
of $10^4$ 
more efficient in the distant past than observed today~\cite{Prantz}. 
A scenario of this sort, if operative in other galaxies as well, 
may also play a desired role on the entropy in 
galaxy clusters and contribute significantly to the extragalactic
$\gamma$ background~\cite{Nath:2005ka}. On the other hand, the extragalactic
$\gamma$ background may have the potential to rule out the required \li6
synthesis by cosmic rays altogether, 
particularly if \li6 has been substantially (factor $10-40$)
depleted in PopII stars. Such large \li6 depletion factors are generally
predicted when \li7 is depleted at a factor of $2.5$.
Finally, a
possible connection between the by WMAP implied fairly early reionisation
of the Universe and the synthesis of \li6 by cosmic rays has also been
considered~\cite{Reeves}.

{\bf Well after the submission of the present manuscript to astro-ph, and
shortly after publication, it was realized that catalytic reactions
involving bound states between the electrically charged stau and nuclei
could considerably change results~\cite{Pospelov,Kohri1}. 
In what follows, to update the paper,
all results of the original version of the manuscript 
are shown in the left panels of figures, whereas results fully accounting
for catalytic effects are shown in the right panels. These are performed by
using the recently computed reaction rates in Ref.~\cite{Hamaguchi,Kamimura}
Since the main
conclusions of the paper are not changed when considering catalytic effects,
the text of the remaining manuscript has not been modified from its 
initial version.
From the figures it is seen that catalytic effects rule out stau decays
with life times exceeding $\tau\approx 10^4$sec due to \li6 overproduction.}

\section{Solving Lithium Problems in the CMSSM}
It may be more economical to suppose that the \li6 was synthesized in
the early Universe, during or right after BBN. This may seem
particularly attractive if the same process also solves the \li7
discrepancy.  In this paper we study this possibility in the
well--defined Constrained Minimal Supersymmetric Standard Model
(CMSSM)~\cite{kkrw} coupled to gravity, under the assumption that the
gravitino is the lightest supersymmetric particle (LSP). In the CMSSM
supersymmetry is broken in a hidden sector and SUSY breaking is
communicated to the visible sector by gravitational interactions. The
CMSSM is parameterised by five quantities: a unified scalar mass
$m_0$, a unified gaugino mass $m_{1/2}$, a unified trilinear coupling
$A_0$ (taken to be zero throughout this paper), $\tanb$ -- the ratio
between the two Higgs vacuum expectation values and $\sgn(\mu)$ where
$\mu$ is the supersymmetric mixing parameter of the two Higgs
doublets.  (We assume $\sgn(\mu)=1$.)  The first of these quantities
are input to the renormalization group equations assuming fixed values
of $\tanb$ and are followed~\cite{djouadi} from the GUT scale to the
electroweak scale in order to determine the supersymmetric mass
spectrum at the weak scale.  Two additional parameters are the
gravitino mass $\mgravitino$ and the reheating temperature $\treh$,
the latter of which is only relevant if TP is efficient.  The relic
density of the NLSP is determined with high accuracy by following the
freezeout from chemical equilibrium. We include all annihilation and
coannihilation channels. For details the reader is referred to
Ref.~\cite{rrc04,Cerdeno:2005eu}.  Whereas a relatively fine scan over
the parameters $m_0$ and $m_{1/2}$ is performed, for the gravitino
mass we assume a number of heuristic relations, i.e. $\mgravitino =
0.2 m_{1/2}$, $\mgravitino = 0.2 m_{0}$, $\mgravitino = m_0$ and
$\mgravitino = 1,10$ and $100\gev$.

For each particular point in the parameter space a complete BBN computation is
performed using the code as introduced in Ref.~\cite{Jedamzik:2004er}. 
This code incorporates
all the relevant hadronic and electromagnetic interactions, including the
most recent data on non--thermal nuclear spallation and fusion reactions, 
required to make precise abundance predictions. A baryonic density of
$\Omega_b h^2 = 0.022$ is assumed leading to the following abundances in the 
absence of particle decay: \li7/H$\approx 4.31\times 10^{-10}$,  
D/H$\approx 2.67\times 10^{-5}$, and \he3/D$\approx 0.39$.
Apart from the relevant accelerator and laboratory limits on particle physics
beyond the standard model each point is also subjected to the following 
observationally inferred limits on the light--element abundances:
$2.2\times 10^{-5}\,\simle\, {\rm D/H} \,\simle\, 5.3\times 10^{-5}$
derived from the D/H abundance in the local interstellar medium and in
high--redshift Lyman--$\alpha$ absorbers, 
${}^3{\rm He/D} \,\simle\, 1.72$ derived from the presolar nebula and,
$Y_p \,\simle\, 0.258$,
where $Y_p$ is the helium mass fraction. 

\bef
\showtwo{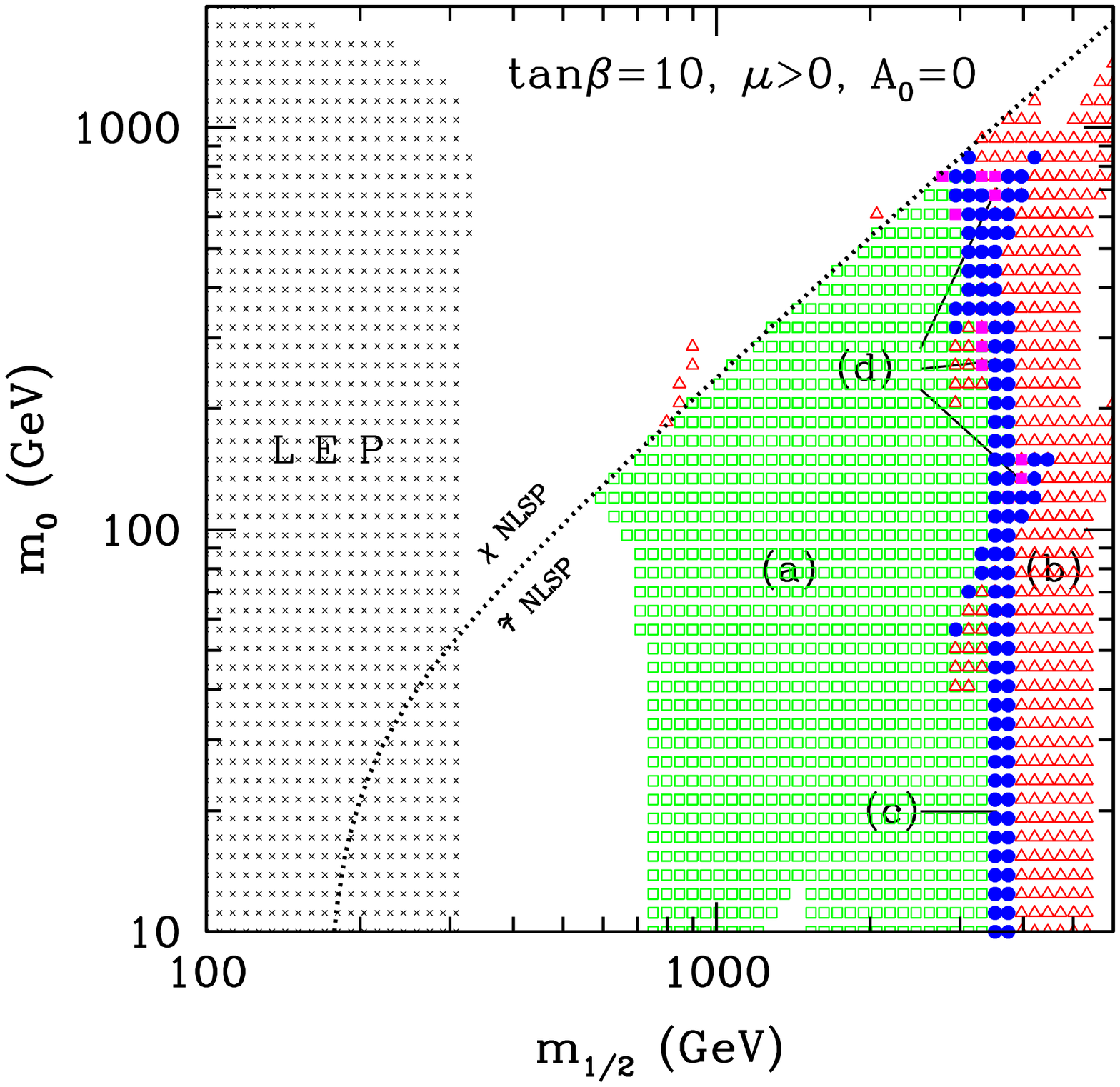}{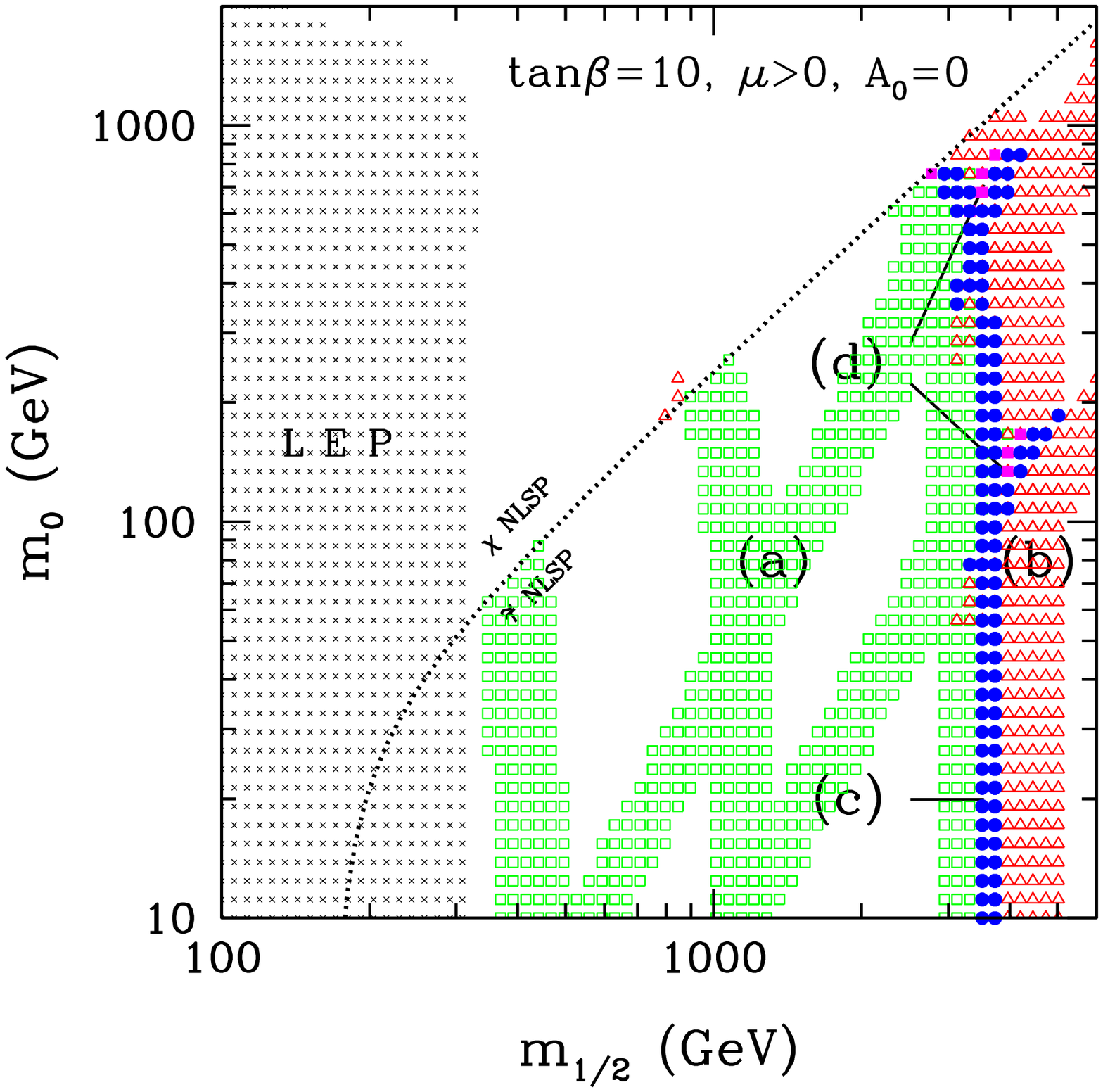}
\caption{Parameter space in the GUT--scale unified supersymmetric
scalar mass $m_0$ -- gaugino mass $m_{1/2}$ plane (all in GeV) where NLSP
decay into gravitinos may resolve one or both of the lithium
problems. The right panel shows results 
with catalytic reactions included, whereas the left panel does neglect 
such reactions.
The parameters of the CMSSM point employed are
$\tanb = 10$, $\mu > 0$, and $A_0 = 0$ and a number of different
gravitino mass $\mgravitino$ choices as explained in the text.
The origin of the \li6 in low--metallicity stars may be
explained (criterium (a), see text) in the area indicated by green
(light grey). The discrepancy between observationally inferred-- and
standard BBN predicted primordial \li7/H abundance may be resolved
(criterium (b), see text)
in the area shown in red (darker grey). Both lithium problems may be
solved at the same time (criterium (c), see text) in the area shown by blue
(darkest grey). When additional stellar \li6 depletion (see text) occurs,
both lithium problems may be resolved
(criterum (d), see text) in the small area shown by pink (grey shading
between the shading of green and red).
}
\label{fig1}
\eef

\bef
\showtwo{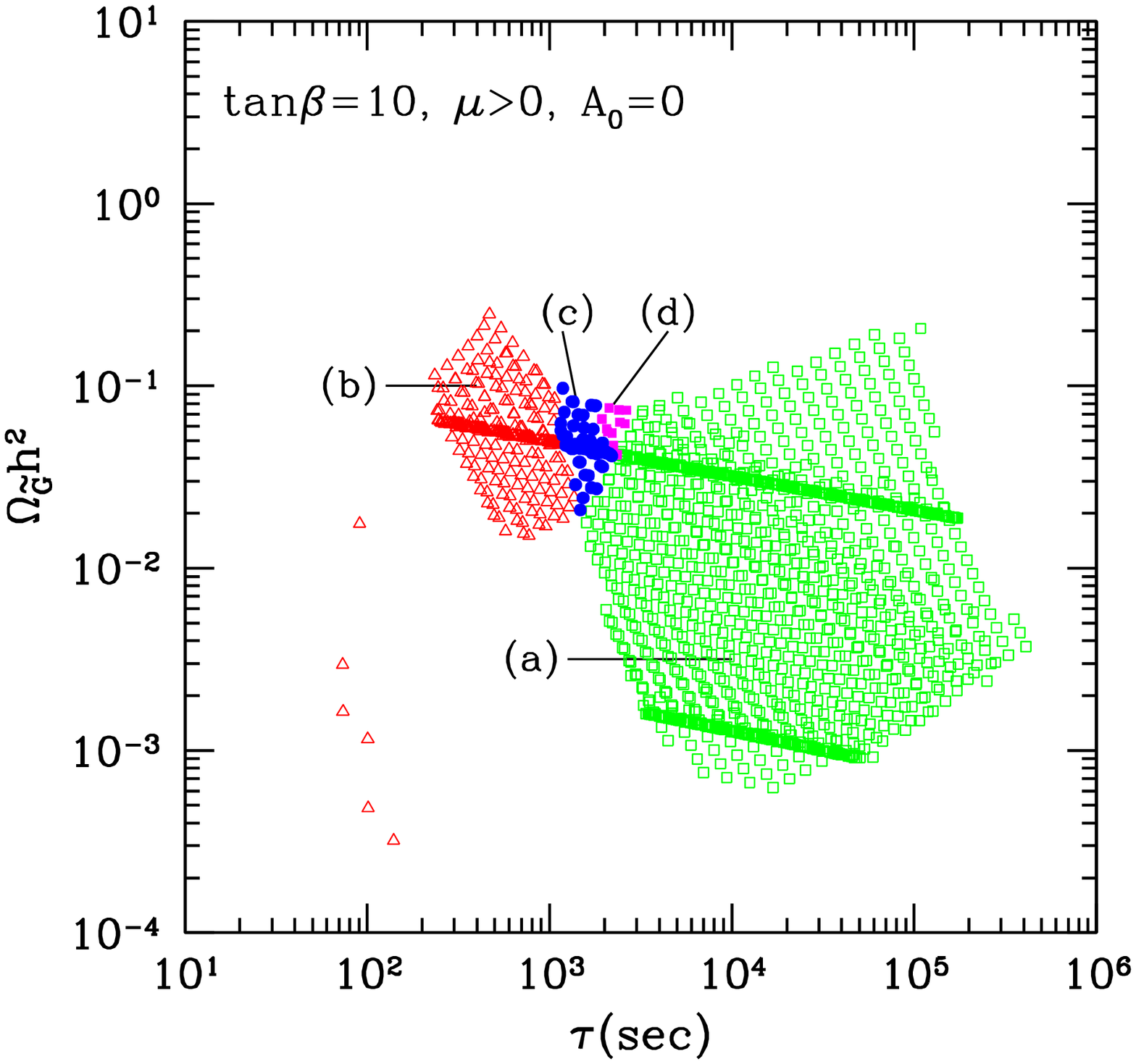}{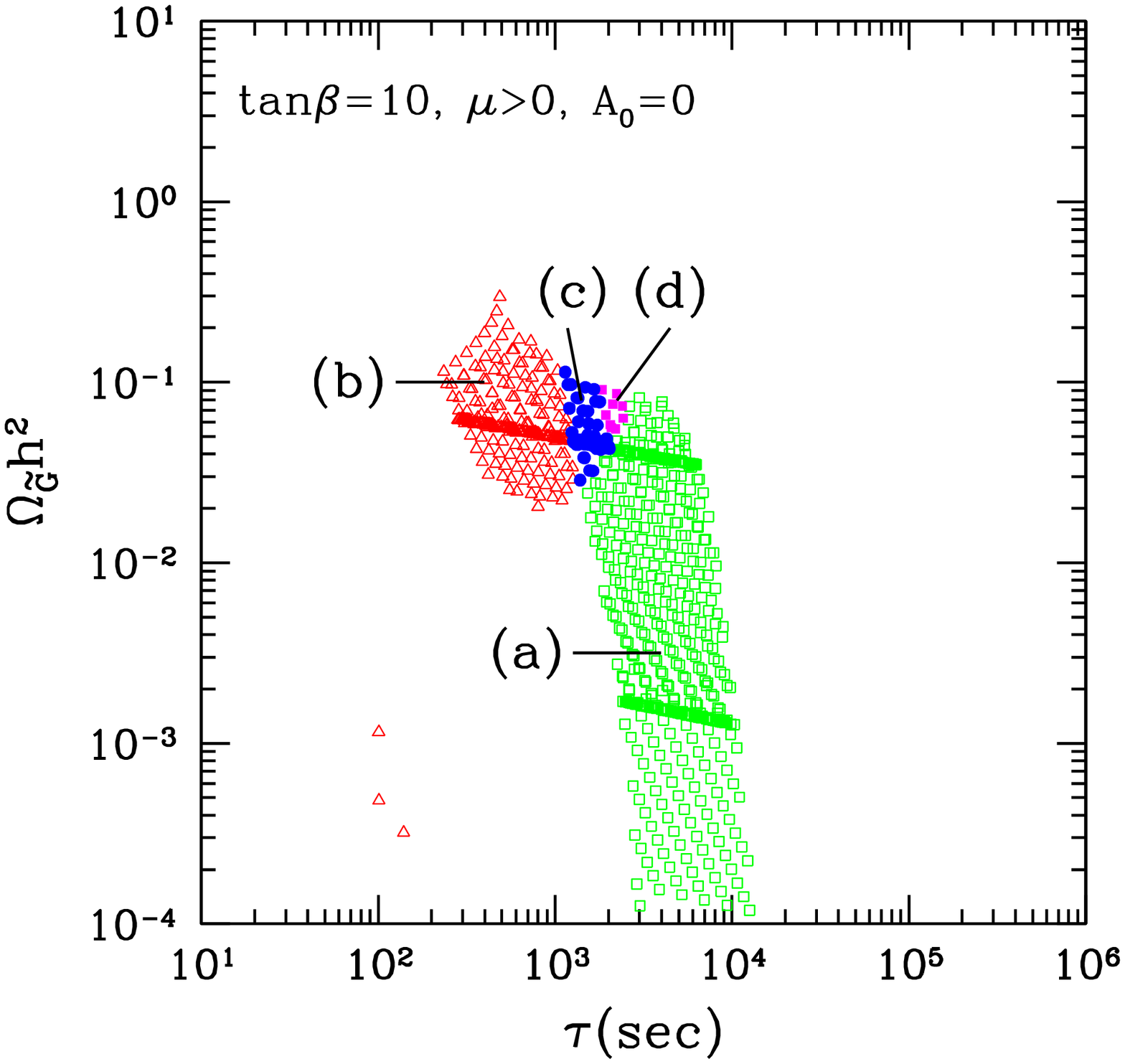}
\caption{Present day gravitino abundance $\Omega_{\tilde{G}}h^2$
as a function of NLSP decay
time in the points shown in Fig.~1. 
The right panel shows results 
with catalytic reactions included, whereas the left panel does neglect 
such reactions.
The color coding is that of Fig.~1.
Here only the gravitino abundance generated during NLSP decay (the
NTP component) is shown. An additional contribution to the gravitino 
abundance could result for a sufficiently high cosmic reheat temperature
$T\sim 10^9$GeV after inflation.
}
\label{fig2}
\eef

\bef
\showtwo{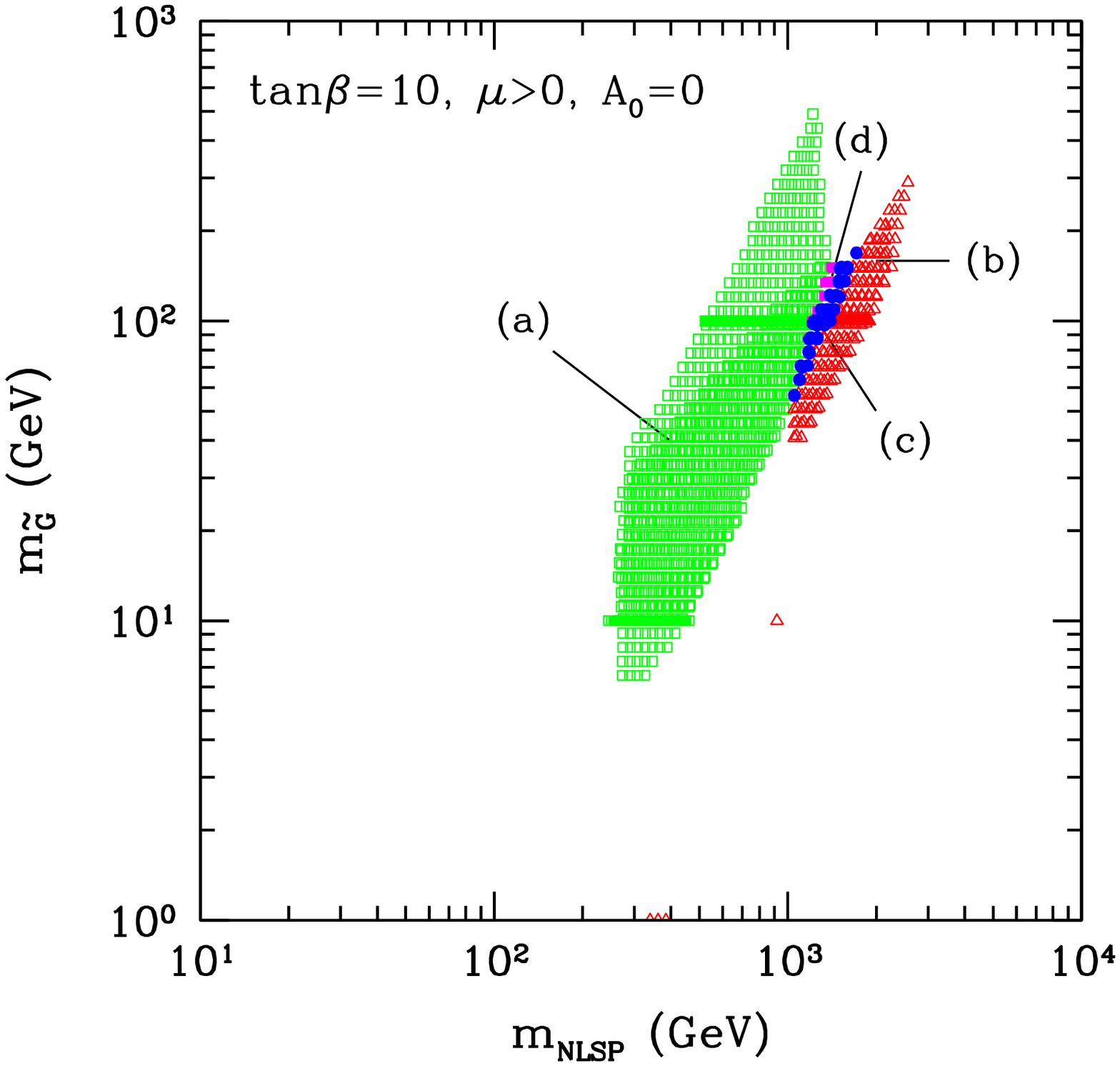}{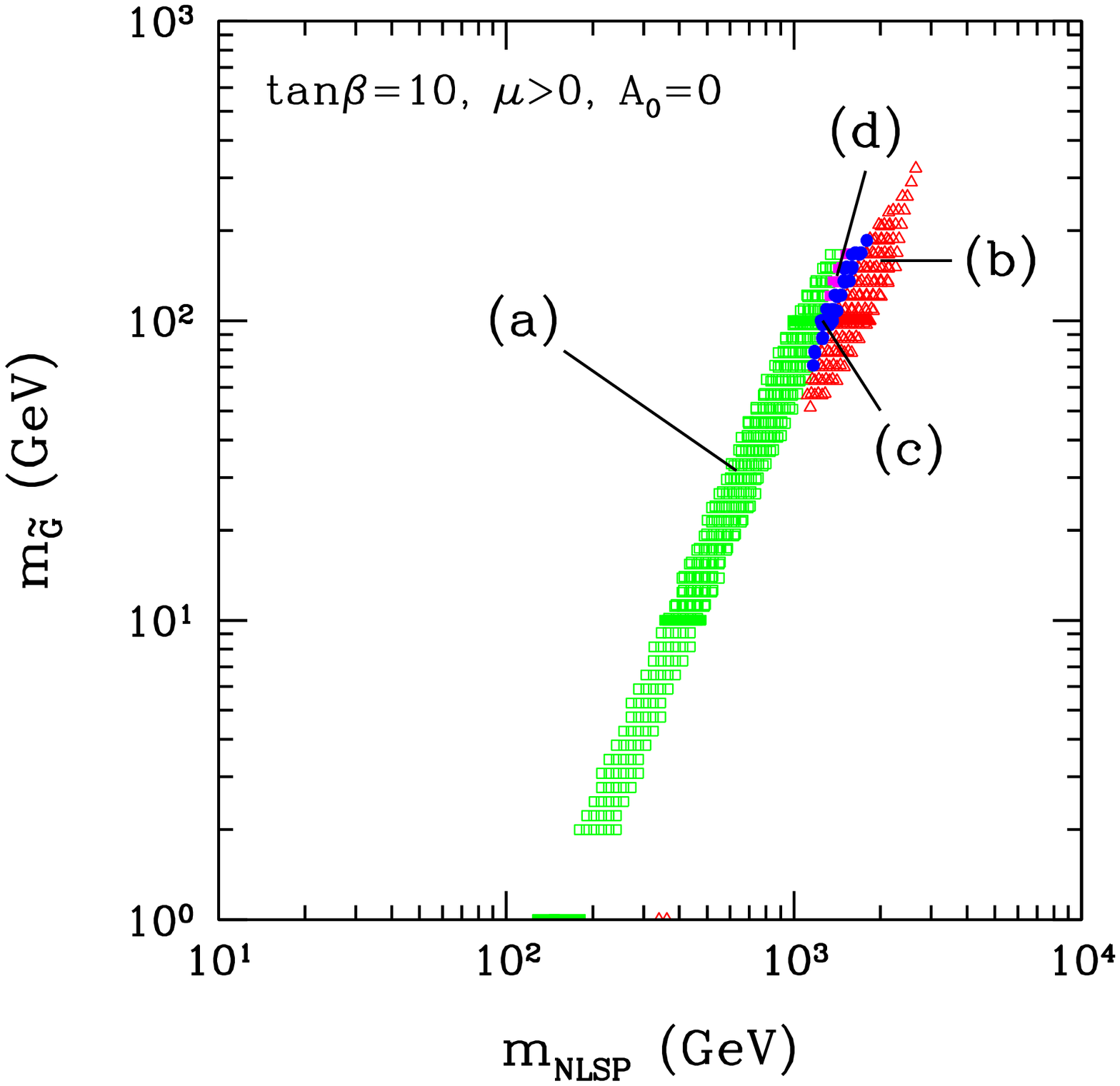}
\caption{Gravitino mass $\mgravitino$ as a function of NLSP mass $m_{\rm NLSP}$
(all in GeV) for those points shown in Fig.~1 and~2.
The right panel shows results 
with catalytic reactions included, whereas the left panel does neglect 
such reactions. 
The color coding is that of Fig.~1.
}
\label{fig3}
\eef

\bef
\showtwo{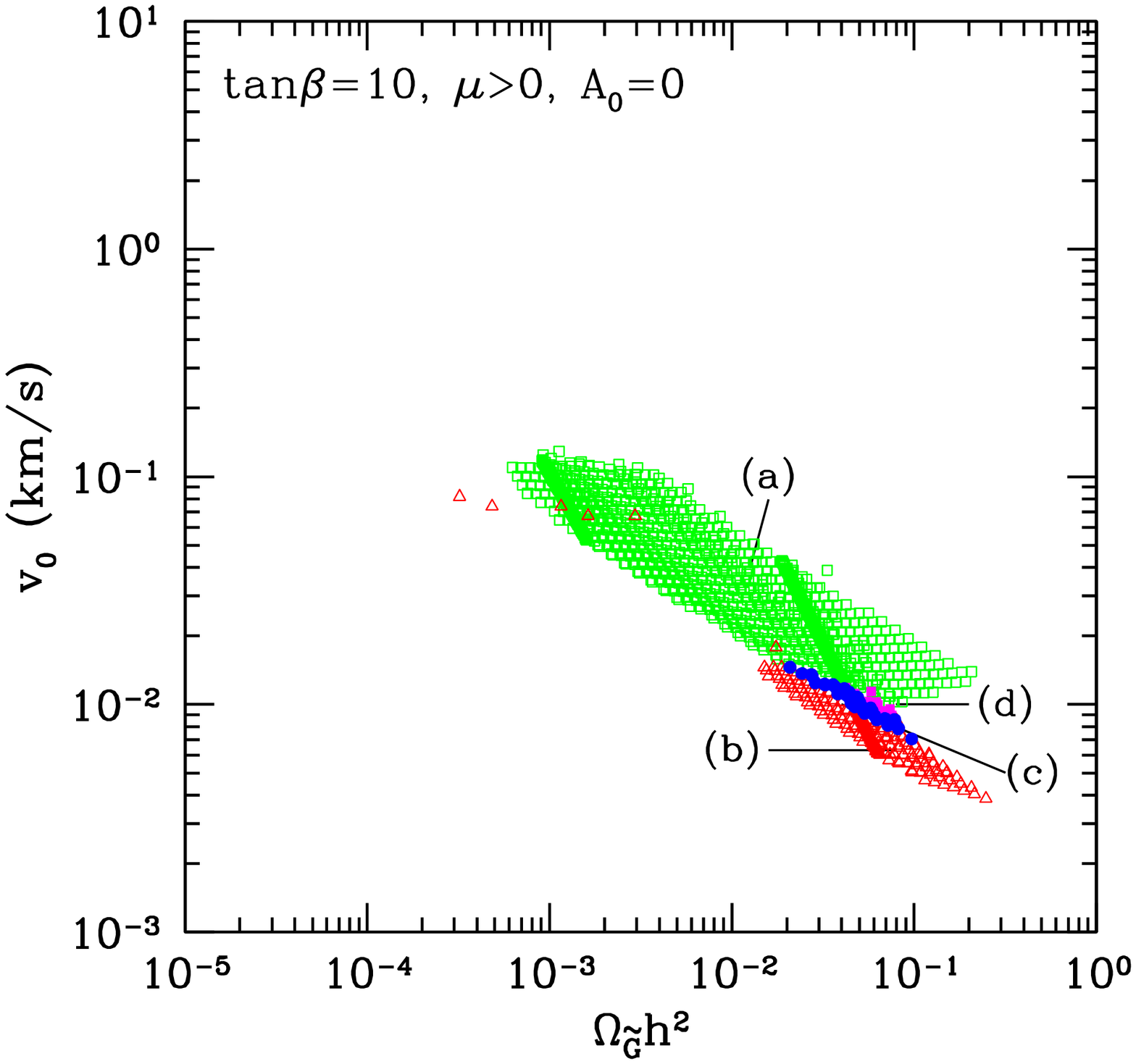}{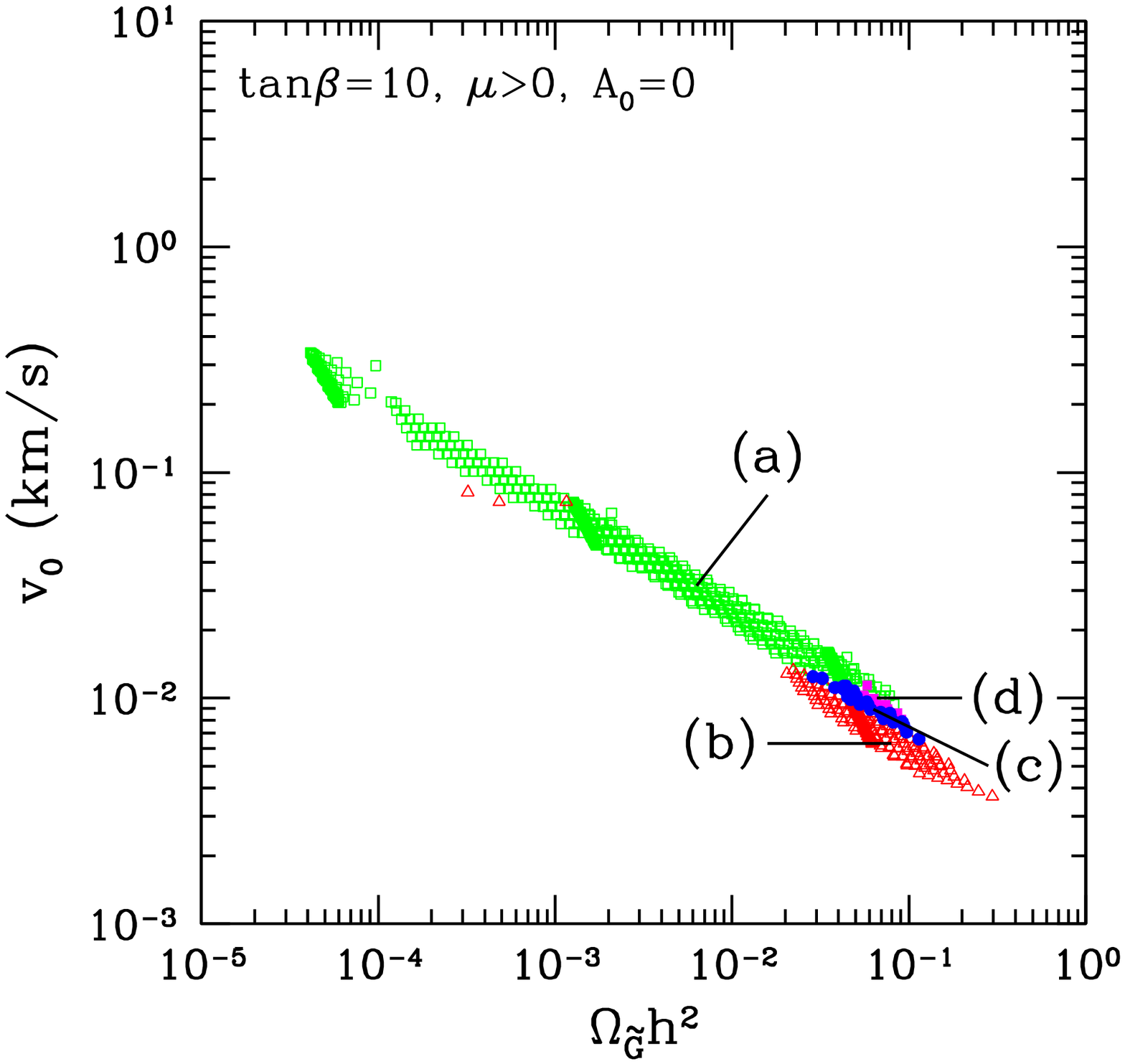}
\caption{Present day free--streaming velocity $v_0$ of the gravitino dark
matter generated during NLSP decay 
as a function of the fractional contribution of gravitinos to the
critical density $\Omega_{\tilde{G}}h^2$
in those points shown in Fig.~1,~2, and~3. 
The right panel shows results 
with catalytic reactions included, whereas the left panel does neglect 
such reactions.
Color coding is
explained in Fig.~1. No reheat--temperature dependent
thermal production of gravitinos has been considered.
}
\label{fig4}
\eef

For the lithium isotopes we
examine three conceptually different possibilities: The \li6 is due to
relic particle decay with the primordial \li7 abundance not much changed.
The discrepancy between the standard BBN predicted and observationally inferred
\li7/H ratio is solved by stellar depletion. A \li7 destruction 
factor of $2.5$ implies generically~\cite{Pins:01,Richard} 
a \li6 destruction of $10-40$. Therefore we apply: 

\vskip 0.1in
\noindent
(a) \li7/H$\,\simge\, 2.5\times 10^{-10}$ and
$0.015\,\simle\,{}^6{\rm Li/{}^7Li} \,\simle\, 3.$ 

\vskip 0.1in
\noindent
The \li7/H ratio is
considerably reduced by relic particle decay with the \li6 due to other
pre--galactic sources. This case corresponds approximately to:

\vskip 0.1in
\noindent
(b) $9\times 10^{-11}\,\simle\,$\li7/H$\,\simle\, 2.5\times 10^{-10}$ 
and \li6/\li7$\,\simle\, 0.015$, 

\vskip 0.1in
\noindent
and finally, both
lithium problems are solved by relic particle decay corresponding to:

\vskip 0.1in
\noindent
(c) $9\times 10^{-11}\,\simle\,$\li7/H$\,\simle\, 2.5\times 10^{-10}$ and
$0.015$$\,\simle\,{}^6{\rm Li/{}^7Li}$$ \,\simle\, 0.15$. 

 As the \li6 isotope is more fragile
than the \li7 isotope, it is conceivable that some stellar 
\li6 depletion has occurred (e.g. on the pre--main--sequence)
even in the absence of \li7 depletion. We have therefore also considered a
case 
\vskip 0.1in
\noindent
(d) $9\times 10^{-11}\,\simle\,$\li7/H$\,\simle\, 2.5\times 10^{-10}$ and
$0.15\,\simle\,{}^6{\rm Li/{}^7Li} \,\simle\, 3.$ 
\vskip 0.1in
\noindent
where the \li6 is in excess
of the observations. This case is shown by pink (with grey--shading between the
shadings of green and red) and is found directly adjacent to the area (c).
We note here that we have relaxed the limit applied in 
Ref.~\cite{Cerdeno:2005eu} on the acceptable \li6/\li7 ratio in order to
account for significant \li6 depletion.

\section{Results}
\vskip 0.1in
Results of our BBN calculations with decaying NLSPs in the CMSSM are shown, 
for the particular case of $\tanb = 10$, $\mu > 0$, and $A_0 = 0$,
in Figs.~1--4 for different choices of $\mgravitino$. 
(We assume low enough $\treh$ so that the TP contribution is negligible.)
Fig.~1 shows the cosmologically interesting
parameter space in the $m_{1/2}$ -- $m_0$ unifying GUT--scale mass
plane.
To help understanding the figures, we remind the reader of some basic
mass relations. The mass of the gluino is roughly given by
$\mgluino\simeq 2.7\mhalf$.  The mass of the lightest neutralino,
which in the CMSSM is almost a pure bino, is
$\mchi\simeq0.4\mhalf$. The lighter stau $\stauone$ is dominated by
$\stauright$ and well above $\mz$ its mass is (neglecting Yukawa
contributions at large $\tanb$) roughly given by $\mstauone^2\simeq
\mzero^2+0.15\mhalf^2$.
Points which satisfy the lithium abundance criteria of case (a), i.e.
providing potentially a solution to the origin of the high \li6 abundance
in low--metallicity stars, are displayed by the green (light grey) points.
It is clear that much of the parameter space may solve the \li6 problem as
this isotope is easily synthesized during a perturbed BBN, either by 
hadronic decays at earlier times $\tau\, \simge\, 10^3$ sec
or by hadronic and electromagnetic decays at 
later times $\tau\, \simge\, 3\times 10^6$ sec. 
If there is any
non--thermal and sufficiently energetic source during or after BBN,
\li6 is normally the first element which is significantly perturbed compared to
the observations~\cite{DEHS}.
The parameter space where only the \li7 abundance is significantly reduced, 
i.e. case (b), is shown by red dots (darker grey). 
This parameter space mostly corresponds
to early decay $\tau \simle 10^3$ sec. Finally, both lithium problems may be
solved at the same time in the area which is shown by blue points (darkest grey).
Except for a degeneracy in $m_0$ this area is well defined, and solutions
are found at $m_{1/2}\approx 3\tev$. It corresponds
exactly to the proposed solution to the lithium problems in 
Ref.~\cite{Jedamzik:2004er},
i.e. the decay of a relic of abundance $\Omega_X h^2 B_h\approx 1-5\times 10^{-4}$
at close to $10^3$ sec after the Big Bang.

The shape of the area which solves both lithium problems (blue)
in Fig.~1 is actually dependent on our discrete
choices of the gravitino mass parameter. Here the vertical band corresponds to
$100\,$GeV gravitinos, whereas the lower- and upper- horizontal bands 
correspond to the choices $\mgravitino =
m_{0}$ and $\mgravitino = 0.2 m_{0}$, respectively. If we were to vary the
gravitino mass as a completely free parameter the blue area would be thus
significantly enlarged.

We alert the reader that due to our consideration of multiple possibilities 
for the gravitino mass, for given $m_0$ and $m_{1/2}$, the parameter space in
Fig.~1 (as well as in Fig.~2--4, see below), shows simultaneously
several choices for the gravitino mass.
This implies that, for example, points which show that criterium (c) is
satisfied
may cover up to coincide
that for the same $m_0$ -- $m_{1/2}$ but a different
$\mgravitino$ also a point satisfying, 
for example, criterium (a) may be potentially found.
We have plotted, from bottom layer to top layer, first all points 
satisfying constraints (a), then (b), (d),
and (c), such that all points satisfying (c) are visible.

It is encouraging that the CMSSM coupled to gravity provides solutions to both lithium
problems. Essentially all of these solutions are obtained in the parameter
space where the stau is the lightest ordinary
supersymmetric partner particle and the NLSP. 
In the CMSSM alone, this parameter space is often claimed to be
cosmologically disfavored due to the electric charge of the stau. 
This is in stark contrast to our findings that points in the 
stau NLSP region should be regarded as cosmologically favored due to
their effect on BBN. Parameter combinations where the bino is the NLSP
are disfavored, when seen in light of the cosmic lithium problems, as the
bino freezeout abundance is usually appreciable 
$\Omega_{\tilde{B}}h^2\sim 1$ and it's hadronic branching ratio is large
due to decay into the $Z$ boson, thus providing too strong
$\Omega_{\tilde{B}}h^2 B_h\gg 5\times 10^{-4}$ of a perturbation to
BBN.

Decay times and final present--day gravitino abundance of the points shown
in Fig.~1 are shown in Fig.~2. Here only the NTP component of gravitinos
produced during NLSP decay is shown, adequate for a low reheat temperature
$\treh$ after inflation.
It is seen that the cosmologically most
appealing points occur indeed for NLSP decay times around $10^3$sec.
Furthermore, it is intriguing to note that those points (blue and pink)
at the same time may provide a significant fraction, or all, of the by
WMAP required dark matter density
$\Omega_{\rm DM}h^2\approx 0.11$ 
in form of the created gravitinos during the decay.  This favorable
coincidence is related to the fact that the hadronic branching ratio
of the stau is usually small $B_h\approx 10^{-4}- 10^{-3}$ since a
higher--order process than the dominant decay into a tau and
gravitino. To obtain $\Omega_{\tilde{\tau}}h^2 B_h\approx 5\times
10^{-4}$ a stau freezeout density of $\Omega_{\tilde{\tau}}h^2\approx
0.5-5$ is required.  Given the stau decay time
$\tau_{\tilde{\tau}}\approx 0.58\,{\rm sec\,} (\mgravitino/1{\rm GeV})^2
(m_{\tilde{\tau}}/1{\rm TeV})^{-5}$ and a typical stau mass of
$1\tev$ the desired decay time $10^3$ sec is obtained for a gravitino
mass of $\mgravitino\approx 50\gev$, i.e. at the electroweak scale.
The gravitino relic abundance $\Omega_{\gravitino}h^2=
\Omega_{\tilde{\tau}}h^2(\mgravitino/m_{\tilde{\tau}})\approx
0.025-0.25$ then comes naturally very close to that required by WMAP. 
This may also be seen in Fig.~3, which shows the stau-- and
gravitino-- masses for the points shown in Fig.~1. The favorite
region (blue and pink) obtains for stau masses around 1 TeV and
gravitino masses between 30 and 200 GeV. Note that in our study in
Ref.~\cite{Cerdeno:2005eu} we had not found regions where the dark
matter abundance due to only the NTP (decay-produced) of gravitinos
may account for the totality of the dark matter, since we had
constrained our analysis to $m_{1/2}\simle 3\,$TeV. Points which only
explain the origin of \li6 are obtained for lighter staus (smaller
$\mhalf$), whereas points which only solve the \li7 discrepancy are
found for heavier staus (larger $\mhalf$).

\bef
\showtwo{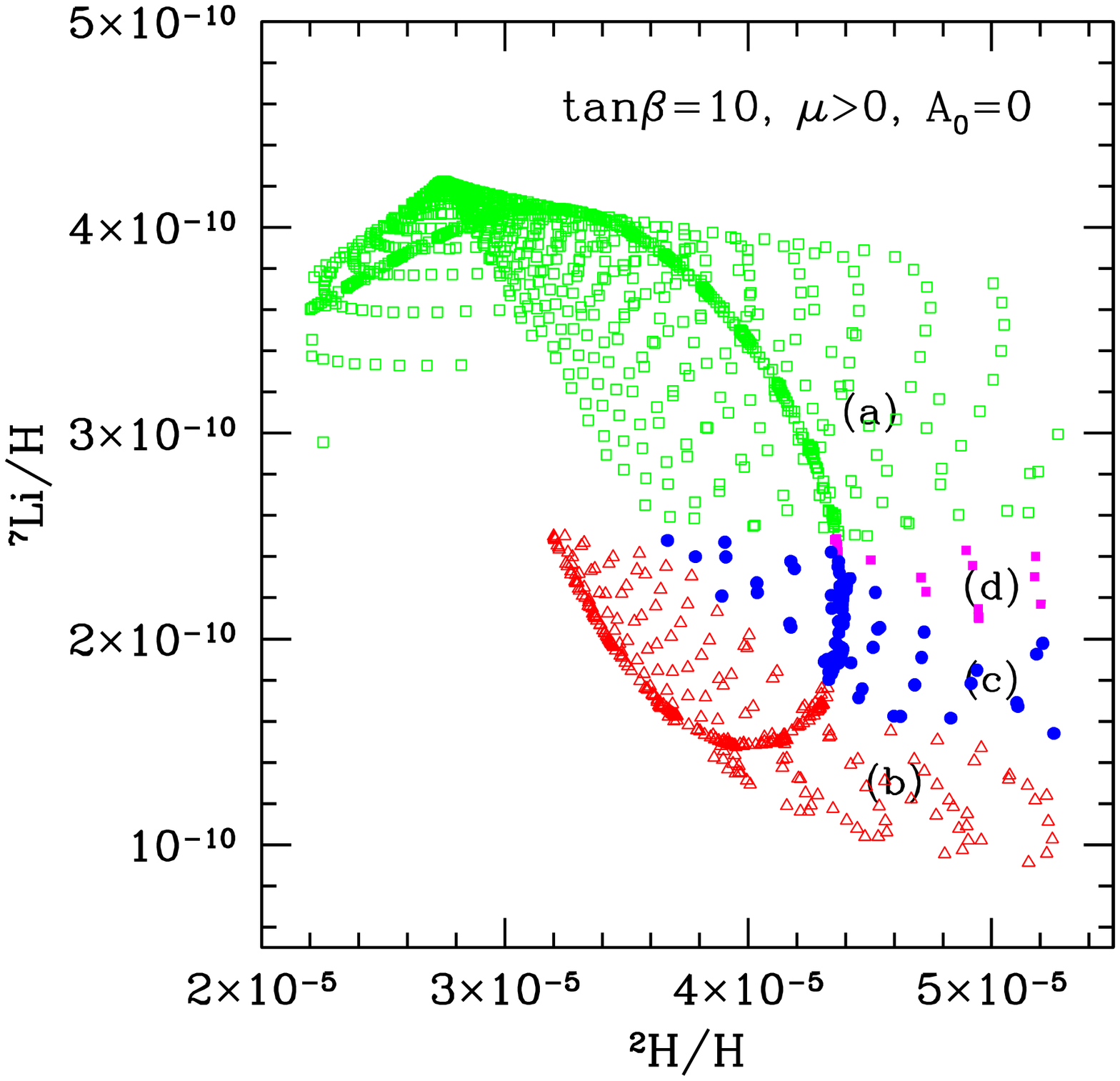}{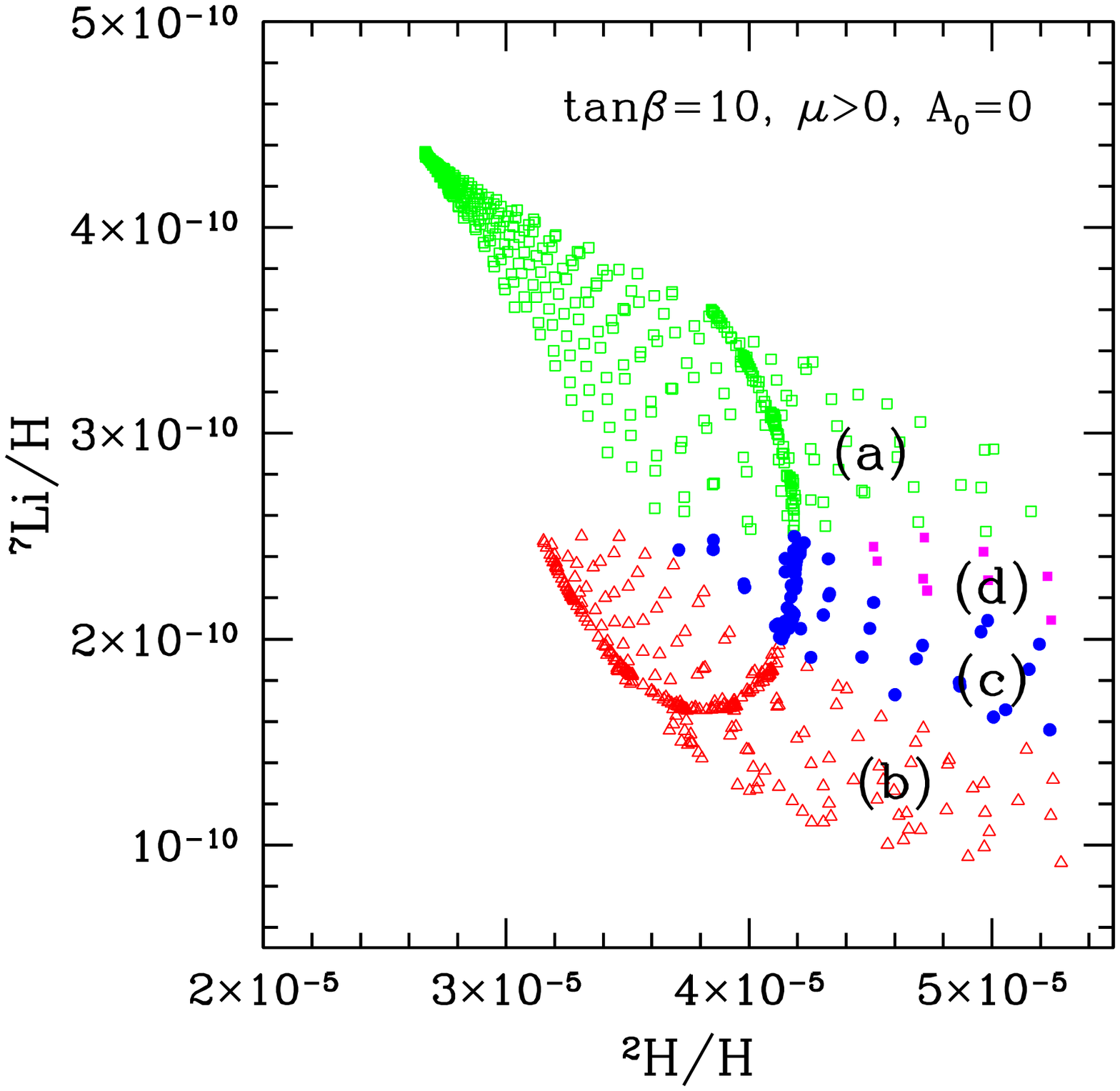}
\caption{BBN yields of \li7/H and \h2/H for the CMSSM parameter space
shown in Fig.1-4. For color coding cf. to Fig. 1.
The right panel shows results 
with catalytic reactions included, whereas the left panel does neglect 
such reactions.}
\label{fig5}
\eef 

\bef
\showtwo{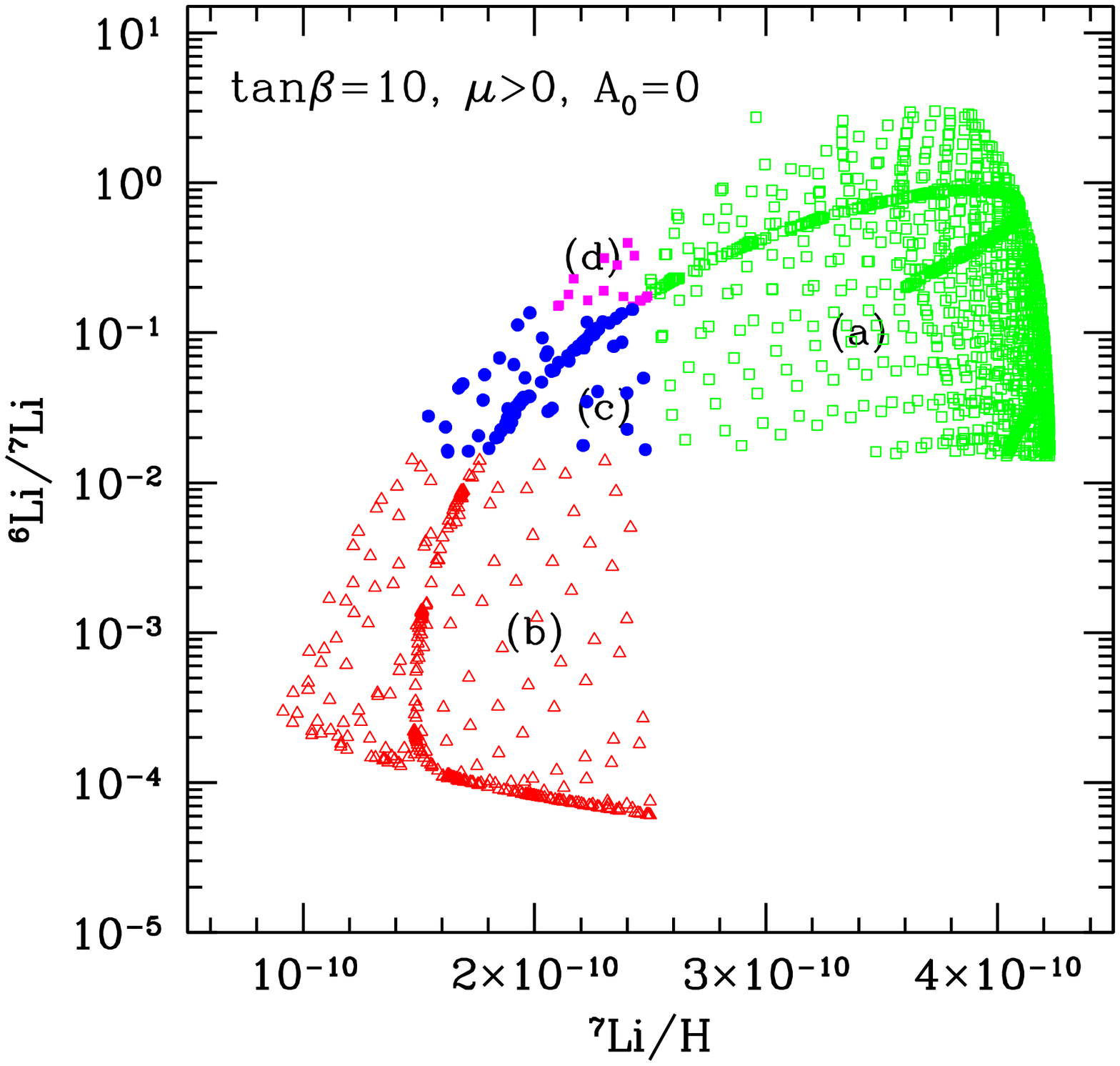}{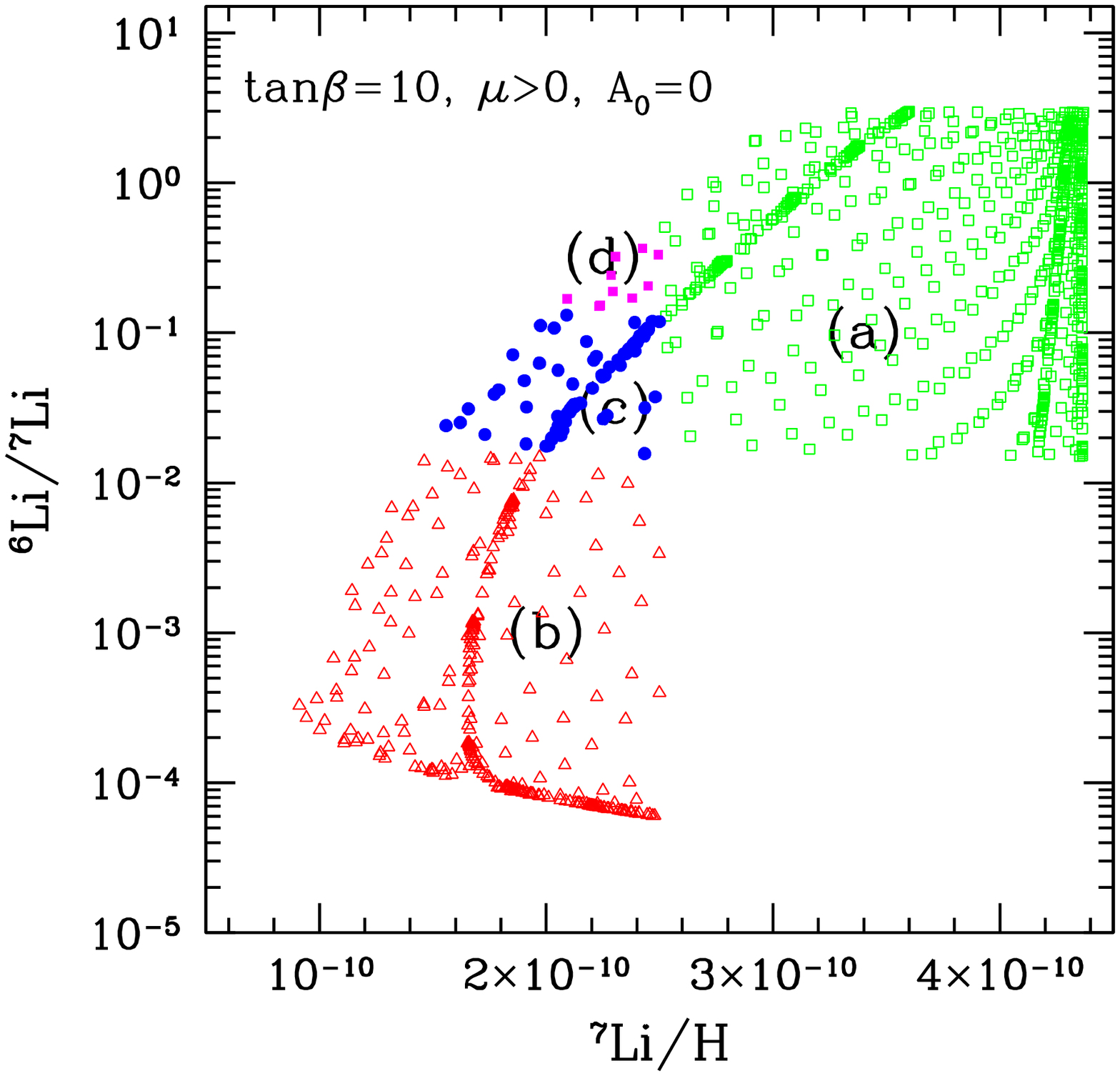}
\caption{BBN yields of \li6/\li7 and \li7/H for the CMSSM parameter space
shown in Fig.1-4. For color coding cf. to Fig. 1.
The right panel shows results 
with catalytic reactions included, whereas the left panel does neglect 
such reactions.}
\label{fig6}
\eef 

We note here that naively one would have expected to find additional solutions
to both lithium problems for decay times around $\tau\approx 10^5$ sec and
correspondingly lighter staus. These solutions could occur since \be7, which
provides 90\% of the primordial \li7 since later electron--capturing, could be
photodisintegrated in a narrow temperature window without the 
photodisintegration of any other element (in particular D). This would be
due to the particular low binding energy of \be7. If this
happened, for the accidentally right hadronic branching ratio 
$B_h\sim 5\times 10^{-5}$ the \be7 could be reduced, and 
\li6 could be produced at the observed level
by the small fraction of hadronic decays. Nevertheless, experimental data
on the photodisintegration process \be7$(\gamma ,\alpha )$\he3 does not
exist. If the cross section for the mirror 
reaction \li7$(\gamma ,\alpha )$\h3~\cite{li7photo} 
is used solutions may be indeed 
found~\cite{remark3}. However, when reverse reaction rate data is used for
the well--studied \he3$(\alpha ,\gamma )$\be7 reaction the \be7
photodisintegration cross section is found abnormally low, such that in
practice, \be7 photodisintegration is always accompanied by some 
observationally unacceptable D photodisintegration.

Gravitino dark matter which is generated by the decay of NLSPs is necessarily
warm(ish), i.e. endowed with free--streaming velocities which impact the
formation of structure in the 
Universe~\cite{Borgani:1996ag,Lin:2000qq,CFRT05,K05,Jedamzik:2005sx}. 
It is known that warm--
and mixed-- dark matter is constrained by a successful small--scale 
structure formation
and a successful early reionisation. Limits from the Lyman--$\alpha$ 
forest~\cite{Lymanwarm}
and the presence of a supermassive black hole at high 
redshift~\cite{Barkana:2001gr} 
are typically
close to $v_{rms,0}\,\simle\, 0.10\,{\rm km/s}$ for the present--day
root--mean--square free--streaming velocity. Potentially even more stringent
limits may be derived by the requirement of early cosmic 
reionisation~\cite{Barkana:2001gr,Yoshida:2003rm}. These may be as strong as 
$v_{rms,0}\,\simle\, 0.03-0.002\,{\rm km/s}$ depending on the exact
reionisation epoch, i.e. $z\sim 17$ as indicated by WMAP or $z\approx 6$ from
high--redshift quasar absorption line systems, as well as on the efficiency
of star formation. On the other hand, warm dark matter has also been claimed
to have beneficial effects on structure formation~\cite{Spergel:1999mh}, 
such as the suppression
of small--scale structure in Milky--Way type halos and the introduction of
constant density cores in dwarf spirals.

\bef
\showtwo{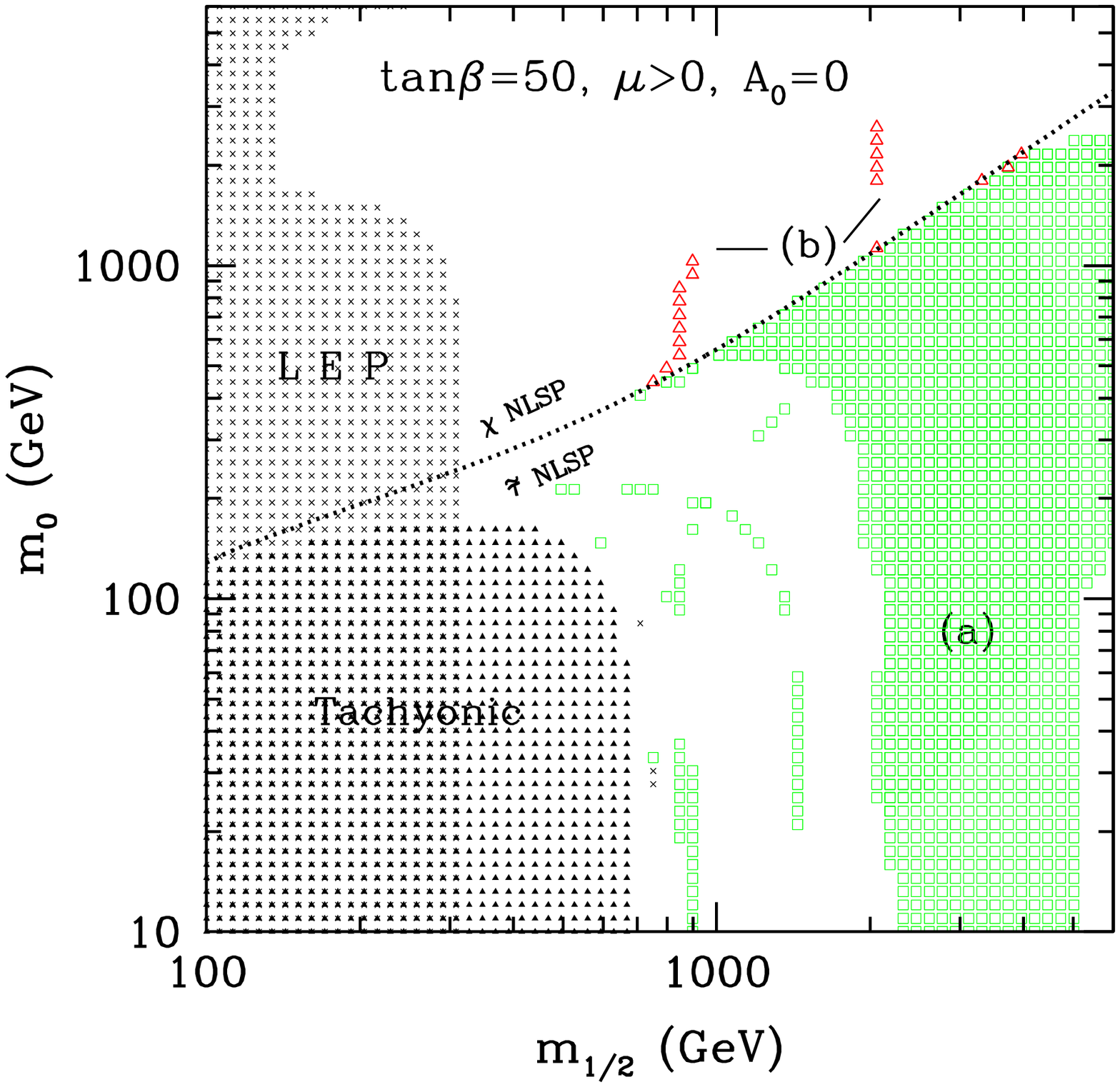}{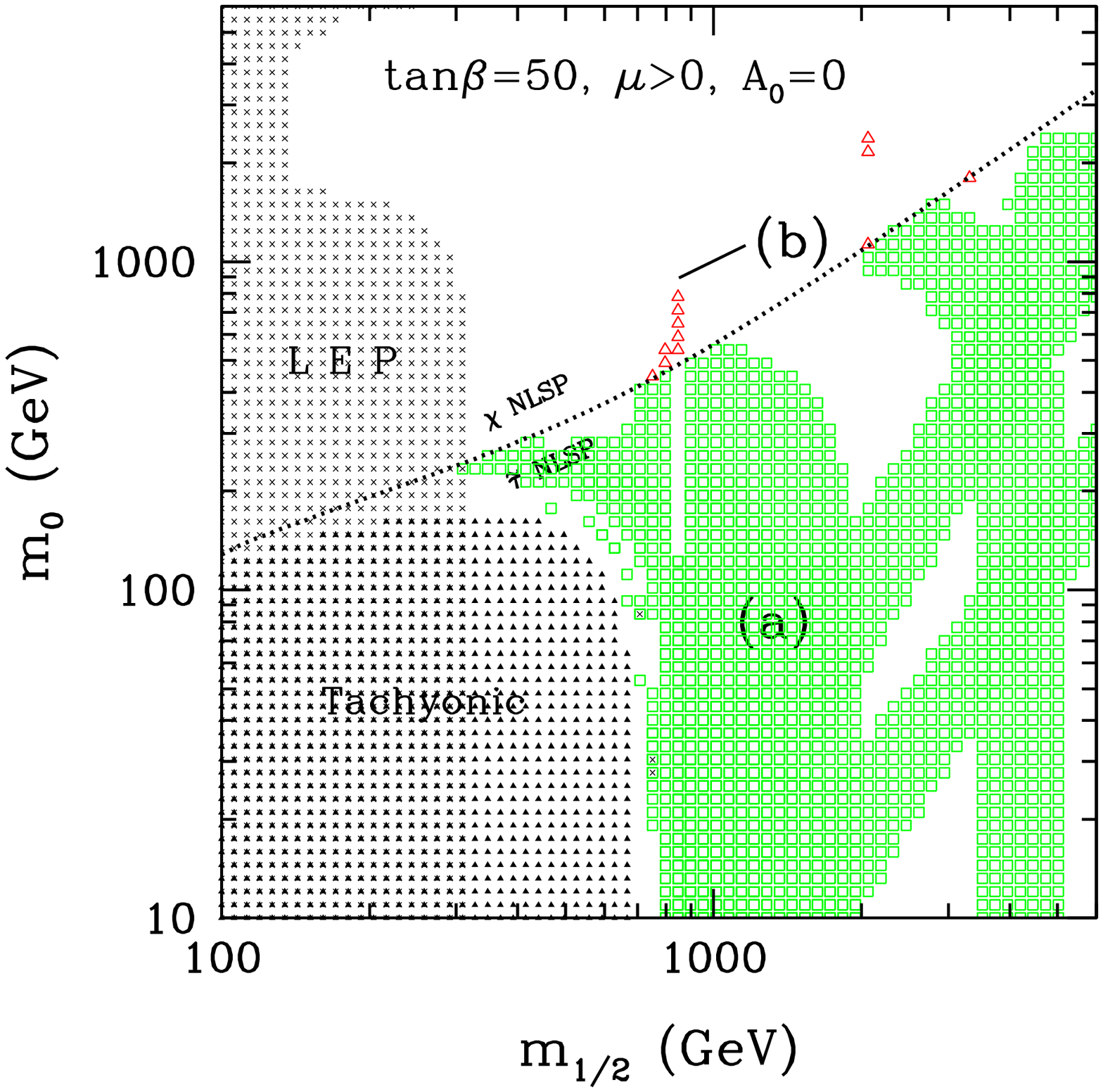}
\caption{As Fig.~1, but for $\tanb = 50$.
The right panel shows results 
with catalytic reactions included, whereas the left panel does neglect 
such reactions.}
\label{fig8}
\eef 

In Fig.~4 we show the present--day free--streaming velocities for the gravitino
dark matter generated during NLSP decay. It is seen that points in the
preferred region which come close to the dark matter density inferred
from cosmological observations
have $v_{rms,0}\,\approx\, 0.007\,{\rm km/s}$. This corresponds to the
free--streaming velocity of a thermally generated gravitino of mass $3.7\kev$,
or equivalently, a reduction of the primordial dark matter power spectrum
compared to that of cold dark matter by a factor two on a scale of $50\,$kpc,
for example.
At present such velocities do not
violate any cosmological constraints, nevertheless, are of a magnitude
which may be interesting to small--scale structure formation. Further
information about the cosmic reionisation history could constrain such points.
Comparatively high free--streaming velocities are reached for points which
satisfy criterium (a). Here the dark matter abundances generated by NLSP
decay are, however, fairly low, such that most of these points are not
ruled out. If accompanied by a component of cold dark matter 
(e.g., TP of gravitinos during reheating) such points
become mixed dark matter.
Constraints on mixed dark matter points from reionisation have
recently been re--analyzed in Ref.~\cite{Jedamzik:2005sx}. 
It is interesting that the
considered scenarios provide an additional verifiable/refutable prediction 
on the warmness of the dark matter. 

BBN abundance yields of D/H, \li7/H, and \li7/\li6 for the points shown
in Figs.1-4 are shown in Figs.~5 and 6. Fig.~5 illustrates that \li7/H
ratios as low as $1.5\times 10^{-10}$ may be synthesized
(corresponding to a \li7 ``depletion'' as much as 0.46$\,$dex) while
producing an observationally satisfying \li6/\li7 ratio as high as
$0.04$. On the other hand, as already noted in 
Ref.~\cite{Jedamzik:2004er} the same scenarios also lead to an increase
in the D/H abundance. The predicted higher
D/H$\simge 3.5\times 10^{-5}$ fares actually
less well with observations~\cite{deut} than the prediction of a SBBN
scenario at the WMAP determined baryon density. However, interpretation of
the available data has to be performed with caution, as there is actually a
dispersion in the inferred D/H ratios in different
low-metallicity Lyman--$\alpha$ absorbers which is
much larger than the inferred errors
in individual D/H determinations. This indicates possibly 
large and unknown
systematic errors as the naive expectation would be to find D/H constant
in different absorbers. 
Furthermore,D/H is essentially always destroyed in stars,
leading to the possibility of a D/H underestimate when much gas has been
cycled through very massive stars. 

We have also considered CMSSM scenarios at $\tanb = 50$,
$\mu > 0$, $A_0 = 0$. Results
for this choice of $\tanb$ are shown in Fig.~1. It is seen that,
though the origin of the \li6 isotope may be explained, there is an
absence of points resolving both lithium problems simultaneously. That is
in contrast to the case $\tanb = 10$.

We have so far concentrated only on gravity--mediated SUSY breaking
where the gravitino mass is expected at the electroweak scale. When
SUSY breaking occurs in a hidden sector which communicates with the
visible sector via gauge--interactions the gravitino is the
LSP and it's mass may be rather small 
$1\,{\rm keV}\,\simle\,\mgravitino\,\simle\, 1\,{\rm GeV}$.
To obtain an NLSP lifetime of $10^3$ sec for, e.g., a
100 MeV gravitino, requires a $90\gev$ NLSP, close to the experimental
bound in case of stau NLSPs. Much lighter gravitinos would require NLSPs
with mass in conflict with LEP data. 
This mass pattern would lead to much larger free--streaming velocities,
i.e. $v_{rms,0}\,\approx\, 1\,{\rm km/s}$. On the other hand such a
case would typically only lead to a very small gravitino dark matter
density due to the small $\mgravitino$. These points are thus not ruled out
and may also solve the lithium problems. However they would 
not provide the bulk of the dark matter. A mass scale of 
$m_{NLSP}\sim 100\,$ GeV implies squarks and gluinos typically
in the several hundreds of GeV up to some 2--3 TeV
range. Supersymmetry at that scale is 
expected to be discoverable at the LHC. In contrast, a solution
of the lithium problems in the CMSSM typically implies squarks and
gluinos in the several TeV range, unlikely to be produced at the
LHC. From the point of view of fine tuning, this range is somewhat
less attractive and for this reason has not been explored
in~\cite{rrc04,Cerdeno:2005eu}. 

Lastly, we mention also that for much of the interesting parameter space
in which the stau is the NLSP, the scalar supersymmetric potential
includes global minima with energy lower than that of the 
Fermi vacuum~\cite{remark6}.
The Fermi vacuum would be thus rendered metastable to decay towards
the global vacuum. 
Since it may, however, be cosmologically sufficiently
long lived, such configurations are not ruled out as long as the Universe
settled into the proper vacuum after inflation. Alternatively, when
$A_0\neq 0$ the Fermi vacuum is often the true vacuum, and such consideration
do not play a role.

\section{Summary}
In conclusion, the observationally inferred primordial \li7 abundance is
a factor $2-3$ lower than that predicted by a standard BBN scenario at the
baryon density as inferred by WMAP. Though it is conceivable that \li7,
which is observed in the atmospheres of low-metallicity stars, has been
destroyed in these stars, there exist currently no self-consistent and
physically motivated scenarios which may explain the observational data. 
In contrast, the \li6 abundance~\cite{asplund} inferred in low--metallicity stars,
an isotope which is usually not associated with BBN but rather
galactic cosmic ray nucleosynthesis, is substantially larger than
those predicted by galactic cosmic ray scenarios. It has been shown that only
under extreme assumptions about putative early cosmic ray 
populations may the \li6 of such a magnitude result~\cite{Prantz}. 
Both, the \li7 and \li6 problems are linked, as significant destruction
of \li7 in low-metallicity stars (a factor of $2.5$) typically
implies an even larger destruction of the more fragile \li6 (a factor
of $10-40$),
making the required synthesis of \li6 by energetic particles even more
problematic.
We investigate here
in the context of the
CMSSM, and under the assumption of the gravitino being the LSP, 
whether one or both of the lithium problems may be solved by 
NLSP decays into gravitinos during or after BBN. Here the NLSP density is
computed self--consistently and with high accuracy.
We have found that there exists ample of supersymmetric parameter space 
where the origin of \li6 may be explained by stau decay during or after
BBN. This had been already shown earlier in the context of hadronic
decays~\cite{DEHS} and electromagnetic decays~\cite{Jeda:00}. 
Similarly, the \li7 may be effectively reduced when staus decay
during BBN. 
It has been shown~\cite{Jedamzik:2004er} that both problems may be solved
simultaneously,
by the decay of a relic particle at about $1000\,$sec after the Big Bang.
These exist for $\tanb = 10$
with 1 TeV staus decaying into $50-200\gev$ gravitinos.
They also lead to an
enhancement of the primordial D/H ratio as compared to that
predicted in SBBN.
By chance, these scenarios result in gravitino abundances
produced during decay, 
$\Omega_{\rm NTP}^{\tilde{G}}$, 
which contribute a large fraction, or all, to the 
by WMAP inferred dark matter
density $\Omega_{\rm DM}$. In cases where 
$\Omega_{\rm NTP}^{\tilde{G}} < \Omega_{DM}$
the gravitino density
may be additionally augmented by production of gravitinos at reheat 
temperatures of $\sim 10^9$GeV. A prediction of these scenarios is the dark
matter to be either lukewarm
(for $\Omega_{\rm NTP}^{\tilde{G}} = \Omega_{DM}$) or mixed
(for $\Omega_{\rm NTP}^{\tilde{G}} < \Omega_{DM}$) 
with free--streaming lengths of a
magnitude interesting to structure formation and relevant to the galactic
core- and substructure problems. Scenarios of this sort are potentially
constrainable/verifiable 
by the process of reionisation, the Lyman-$\alpha$ forest, and/or
weak lensing.
It would indeed be interesting if the
anomalies inferred in the abundances of the lithium isotopes would 
present us with information about the
nature of the dark matter and physics beyond the standard model.

\vspace*{0.5cm}
\noindent
{\bf Acknowledgements}\\
We acknowledge discussions with 
Martin Asplund, Marina Chadeyeva, Pierre Descouvemont, 
Jean-Loic Kneur, Gilbert Moultaka, 
Nikos Prantzos, Olivier Richard, and
Gary Steigman.
K.--Y.~C. are funded by PPARC. We further acknowledge funding
from EU FP6 programme -- ILIAS (ENTApP).\\


\begin{thebibliography}{99}
\bibitem{remark1}
The only assumption here is, as in the case of neutralinos, that the
Universe is in thermal equilibrium at the electroweak scale.
\bibitem{remark2}
for earlier works on relic particle decay the reader is
referred to Refs.~\cite{RS} and ~\cite{DEHS}.
\bibitem{remark3} There are, however, also large mutual inconsistencies
in the published \li7$(\gamma ,\alpha )$\h3~\cite{li7photo} photodisintegration
cross section.
\end{thebibliography}
\end{document}